# Deep Learning Empowered Sub-Diffraction Terahertz Backpropagation Single-Pixel Imaging

*AUTHOR NAMES. Yongsheng Zhu, Shaojing Liu, Ximiao Wang, Runli Li, Haili Yang, Jiali Wang, Hongjia Zhu, Yanlin Ke, Ningsheng Xu, Huanjun Chen*, Shaozhi Deng**

State Key Laboratory of Optoelectronic Materials and Technologies, Guangdong Province Key Laboratory of Display Material and Technology, School of Electronics and Information Technology, Sun Yat-sen University, Guangzhou 510275, China.

*Corresponding authors: chenhj8@mail.sysu.edu.cn, stsdsz@mail.sysu.edu.cn

**ABSTRACT.** Terahertz single-pixel imaging (THz SPI) has garnered widespread attention for its potential to overcome challenges associated with THz focal plane arrays. However, the inherently long wavelength of THz waves limits imaging resolution, while achieving subwavelength resolution requires harsh experimental conditions and time-consuming processes. Here, we propose a sub-diffraction THz backpropagation SPI technique. We illuminate the object with continuous-wave 0.36-THz radiation ($\lambda_0$ = 833.3 μm). The transmitted THz wave is modulated by prearranged patterns generated on a 500-μm-thick silicon wafer and subsequently recorded by a far-field single-pixel detector. An untrained neural network constrained with the physical SPI process iteratively reconstructs the THz images with an ultralow sampling ratio of 1.5625%, significantly reducing the long sampling times. To further suppress the THz diffraction-field effects, a backpropagation SPI from near field to far field is implemented by integrating with a THz physical propagation model into the output layer of the network. Notably, using the thick wafer where THz evanescent field cannot be fully recorded, we achieve a spatial resolution of 118 μm (~$\lambda_0$/7) through backpropagation SPI, thus eliminating the need for ultrathin photomodulators. This approach provides an efficient solution for advancing THz microscopic imaging and addressing other inverse imaging challenges.

**KEYWORDS.** terahertz imaging, sub-diffraction imaging, single-pixel imaging, neural network, computational imaging

## INTRODUCTION

Terahertz (THz) waves (0.1–10 THz), characterized by their non-ionizing photon energy, are transparent to many non-metallic and non-polar materials, making them highly promising for biomedical imaging,[1,2] safety inspection,[3] and non-destructive testing.[4,5] However, due to diffraction effects and the relatively large THz wavelength, achieving micron-level imaging resolution remains challenging. Subwavelength THz imaging currently relies mainly on near-field raster scanning techniques,[6] using sub-diffraction-limited THz atomic force microscope[7] or scanning tunneling microscope tips.[8] These methods record the evanescent field through pixel-by-pixel sampling on the sample surface, achieving micron- or even atomic-level resolution. Despite significant advancements in THz near-field probe techniques, the weak signals emitted by these probes are highly sensitive to detector noise, often requiring lengthy measurement times.

Recent advances in THz wave modulation using spatial light modulators have played a pivotal role,[9,10] particularly in enabling THz imaging methods using spatial encoding masks in conjunction with a single-pixel detector, known as ghost imaging (GI)[11-13] or single-pixel imaging (SPI).[14-18] SPI techniques offer significant potential to overcome the challenges associated with time-consuming raster scanning,[19-21] as well as the high cost THz focal plane array. Recently, the SPI technique has been demonstrated for far-field sub-diffraction THz imaging, enabling the recovery of sub-diffraction spatial information hidden within diffracted fields by analyzing the intensities recorded by a single-pixel detector. The achieved spatial resolutions range from $\lambda_0/2$ to $\lambda_0/100$.[22-29] This is typically done using Hadamard SPI (HSPI)[17] and compressed sensing (CS) algorithm,[22,24] where the object is first encoded with patterned THz waves in the near field. A single-pixel detector in the far field then collects the total intensity of the encoded THz image, and post-processing of the detected THz scattering field signals associated with the deterministic patterns enables the reconstruction of the near-field THz image. For that end, ultrathin THz photomodulators, such as 6-μm-thick silicon wafers[23,24] or 180-nm-thick $VO_2$ films,[26] are commonly used to access the evanescent field around the target. Another approach involves structured near-filed THz detection or generation through the photoexcitation of femtosecond laser pulses.[27,30,31] However, in practical subwavelength imaging, objects are rarely infinitesimally thin, and the diffraction effects arising from the propagation distance between the THz source plane and the thick object lead to a reduced in spatial resolution. While time-resolved nonlinear GI[25,28,29] has demonstrated that combining standard SPI with a backpropagation operation under a high sampling ratio (*SR*) can recover higher-resolution near-field THz images, the use of undersampling strategies often results in ill-posed problems in THz image reconstruction. To achieve high-quality imaging with reduced near-field sampling, an alternative approach is deep learning-based THz SPI techniques employing supervised training strategies.[32] While these approaches have achieved far-field THz image reconstruction with *SR*s below 10%, they require training deep neural networks on extensive datasets, resulting in time-consuming training processes and limited generalization capabilities.[33-35] Recently, an untrained neural network has been employed to address the inverse problem in computational imaging.[32,36] Unlike supervised neural networks, which require large labeled data to optimize their weights and biases, the untrained neural network operates without pre-training, thus eliminating the need for tens of thousands of input−output data pairs typically required to establish a robust mapping relationship in supervised learning. Similar concepts have been applied in computational imaging fields such as phase retrieval,[37] super-resolution imaging,[38,39] and THz holographic imaging.[40,41] However, applying this network to THz SPI, particularly for effective integration of near-field THz characteristics with backpropagation under undersampling, remains unexplored.

In this work, we propose a sub-diffraction THz backpropagation SPI framework using a continuous-wave (CW) monochromatic THz source. Our framework structurally encodes the THz diffraction image using a passivated silicon photomodulator, and a THz detector records the field intensity information, facilitating comprehensive data acquisition regarding the interactions between THz waves and objects. The one-dimensional vector from the detector, associated with the coding patterns, serves as the sole input for an untrained neural network to perform THz SPI. Compared to state-of-the-art THz SPI methods, the proposed framework, termed untrained THz SPI based on deep learning (untrained TSPIDL), significantly enhances the quality of the THz image at a much lower *SR*. To the best of our knowledge, this is the first application of an untrained neural network to THz SPI. We further model the forward-propagation process of the THz diffraction field and constrain the output of the neural network with the angular spectrum propagation (ASP) model. As the network undergoes iterative optimization, it is ultimately compelled to output a clear THz image. The application of this principle to binary amplitude objects has been demonstrated, achieving a near-field resolution of $\lambda_0/7$ at an ultralow sampling ratio of 3.125%, and demonstrating the ability to reconstruct from near-field to far-field across different silicon planes. This technique not only significantly reduces sampling time but also provides improved sub-diffraction resolution with the same thickness of photomodulators, while maintaining THz sensing for non-invasive detection properties.

## METHODS

The principle of THz SPI involves patterning a THz image using time-varying optical patterns, with a single-pixel detector recording the transmitted or reflected THz signals. The detector readout is then combined with the corresponding spatial patterns to reconstruct the THz images. The impact of the distance $d$ between the object plane $z = 0$ and the recording plane $z = d$, such as a silicon wafer in our current study, on imaging remains an open question. After the THz wave impacts the object and propagates to the back surface of the silicon plane, it no longer maintains the original spatial distribution but instead takes on a diffracted form. This diffraction unavoidably blurs the image reconstruction of subwavelength objects.

As illustrated in Figure 1, the principle of our proposed untrained TSPIDL is based on THz wave fluctuations, which strictly rely on diffraction and automatically perform spatial integration to provide the necessary operational conditions for imaging. Assuming a THz plane wave with a wavelength of $\lambda$ is normally incident on the object plane $z = 0$, the complex amplitude of the transmitted THz field can be expressed as:

$$E_0\left(x_0, y_0; z=0\right) = A_0(x_0, y_0)e^{i\varphi_0(x_0, y_0)} \tag{1}$$

where $A_0$ and $\varphi_0$ are the amplitude and the initialized phase at the object plane $z = 0$. For the forward propagation distance $d = d_{FP}$ from the object plane $z = 0$ to the recording plane $z = d$, the complex amplitude $E_d$ can be obtained by the ASP theory, and the scalar diffraction integral formulas are represented as:[23,42,43]

$$E_h(x,y)=\iint_{u_x^2+v_y^2\le 1} E_0(f_x,f_y)e^{i2\pi(f_x x_0+f_y y_0)}e^{jkd\sqrt{1-u_x^2-v_y^2}}df_x df_y \tag{2}$$

$$E_e(x,y)=\iint_{u_x^2+v_y^2>1} E_0(f_x,f_y)e^{i2\pi(f_x x_0+f_y y_0)}e^{-kd\sqrt{u_x^2+v_y^2-1}}df_x df_y \tag{3}$$

where $\lambda$ is the THz wavelength, $k=2\pi/\lambda$ is the free space wavenumber, $d$ is the propagation distance, $E_0(f_x,f_y)$ is a spectral amplitude function that is the Fourier transform of the object field distribution $E_0(x_0,y_0)$ in the plane $z = 0$ with the spatial frequencies $f_x = x/\lambda d$ and $f_y = y/\lambda d$ in the $x$ and $y$ directions, $u_x=\lambda f_x$ and $v_y=\lambda f_y$ are normalized spatial frequencies. The THz field $E_d(x,y)$ is the sum of the homogeneous propagating field component $E_h(x,y)$ and the evanescent field component $E_e(x,y)$. From eq 3, it is evident that as the propagation distance increases, the evanescent fields with higher spatial frequencies attenuate more rapidly.

The diffraction image $O_d(x,y)=|E_d(x,y;z=d)|^2$ is encoded by a series of spatial patterns (scheme shown in left part of Figure 1), and the detector readout is denoted as:

$$I_i=\iint O_d(x,y)P_i(x,y)dxdy+\eta \tag{4}$$

where $P_i$ and $I_i$ represent the spatial response of the $i_{th}$ pattern and the detector readout of the $i_{th}$ measurement, respectively. $\eta$ denotes the background noise. For patterns $P = \{P_i (x, y) \mid i = 1, 2, 3, \ldots, M\}$ with a length of $M$, the encoding process can be expressed as a matrix convolution operation, resulting in a one-dimensional (1D) vector $I = \{I_i \mid i = 1, 2, \ldots, M\}$ that is used as the direct input to the untrained neural network. Due to the ill-posed reconstruction caused by undersampling strategies, appropriate prior assumptions are necessary to compensate for the missing information.[44] Therefore, to enhance the noise robustness, an intermediate physical encoding layer is constructed as an indirect input to the network. For $I$ and $P$, a differential ghost imaging (DGI) algorithm[45] based on second-order correlation is used to compute a rough estimate of the diffracted THz image $\hat{O}'_d(x,y) = g^{(2)}(I;P)$ on the recording plane, represented as:

$$\hat{O}'_d(x,y) = \langle (P_i(x,y) - \langle P_i(x,y)\rangle_M) \times (I'_i(x,y) - \langle I'_i(x,y)\rangle_M \rangle_M \tag{5}$$

where $\langle \cdot \rangle_M$ denotes the ensemble average approximately defined as $\langle P_i(x,y)\rangle_M = \frac{1}{M}\sum_{i=1}^{M} P_i(x,y)$.

$I'_i(x,y) = I_i - \frac{\langle I_i \rangle_M}{\langle S_i \rangle_M} S_i$ denotes the normalized differential signal, and $S_i = \sum_{x,y} P_i(x,y)$ is the total intensity distribution of $P$. By mapping the physical features in the measurement space to the preliminarily estimated THz diffraction image space, an interpretable prior extraction layer is provided without the need for complex computations. The output $\hat{O}_0(x_0,y_0) = \left|\hat{E}_0\left(x_0,y_0\right)\right|^2$ of the network is an estimated intensity distribution of the object field, represented as:

$$\hat{O}_0(x_0,y_0) = f_\theta(g^{(2)}(I;P)) \tag{6}$$

where $f_\theta$ represents the mapping function of the neural network with the weight and bias parameters $\theta$. The implementation of $f_\theta$ is not restricted to any specific network architecture. However, considering the priorities of computational efficiency and minimizing reconstruction time, we implemented a lightweight model comprising four main types of convolutional blocks (Figure S1, Supporting Information).

The intensity distribution $\hat{O}_d(x,y) = \left|\hat{E}_d\left(x,y\right)\right|^2$ is obtained by substituting the backpropagation distance $d = d_{BP}$ into eq 2 and eq 3. This distribution is subsequently is convolved with the spatial patterns $P_i(x, y)$ to yield the 1D intensity estimate $\hat{I}_i$, which is calculated as:

$$\hat{I}_i = \hat{O}_d(x,y) \otimes P_i\left(x,y\right) \tag{7}$$

The mean squared error (MSE) between the 1D intensity estimated $\hat{I}$ by eq 7 and the experimentally measured value $I$ from eq 4 is minimized by measuring the Euclidean distance between the predicted output and the desired input (MSE loss shown in Figure S2 of Supporting Information). The weights and biases of the network are optimized through gradient descent, and the loss function of the network is expressed as:

$$f_{\theta^*} = \underset{\theta}{\operatorname{argmin}} \left\| I - \hat{I} \right\|^2 + \mathrm{TV}(\hat{O}_0(x_0, y_0)) \tag{8}$$

where $f_{\theta^*}$ is the mapping function of the neural network in the training process, TV is regularization term used to avoid falling into local minimization of the untrained TSPIDL and improve the reconstruction results.[36,38] This process does not require pre-training on a large labeled dataset. Instead, it directly operates on an unsupervised learning strategy through iterative optimization constrained by imaging physics principles, thereby yielding a generalizable model.

Specifically, the spatial resolutions $u_x$ and $u_y$ can be calculated based on the recording plane dimensions and pixel resolution. These values, along with the backpropagation distance $d = d_{BP}$ (which equals the forward propagation distance $d = d_{FP}$ of the actual imaging system), are substituted into eq 2 and eq 3. Following this, the untrained TSPIDL model takes $I$ as input and, utilizing eq 2–7, generates $\hat{I}$ in each iteration. The parameters $\theta^*$ are subsequently updated through the loss function defined in eq 8. As the iteration process continues, the estimated output $\hat{I}$ is constrained to converge to the actual measured values $I$ from the detector. As shown in eq 6, the process of iterative optimization enables the network $f_{\theta*}$ to indirectly establish the mapping relationship between the experimental measurements $I$ and the object-field intensity distribution $\hat{O}_0(x_0, y_0) = f_{\theta*}(g^{(2)}(\hat{I}; P))$ without requiring pre-trained parameters from a large labeled dataset. Thus, a backpropagation process is achieved by retrieving the backpropagation distance $d = d_{BP}$, which refocuses the blurred THz diffracted image from the defocused recording plane $z = d$ back to the object plane $z = 0$. This method is applicable not only for reconstructing sub-diffraction resolution when the object is in close contact with the modulator plane but also in scenarios where the object is situated beyond the THz near-field region. At an arbitrary distance outside the modulator plane, the refocusing process resembles that of a standard diffraction-limit imaging system.

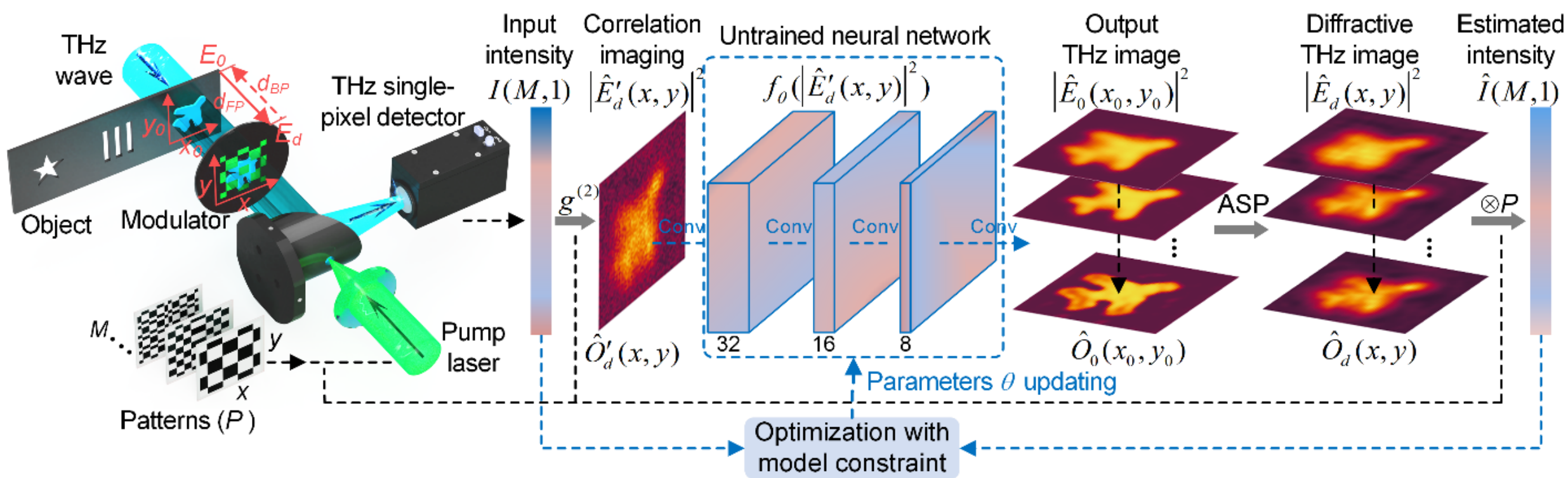

**Figure 1.** Schematic illustration of the sub-diffraction THz backpropagation SPI pipeline. Left panel: A THz wave propagates from the object plane $E_0$ to the photomodulator back surface $E_d$, generating a THz diffraction field at distance $d = d_{FP}$. The modulator back surface is simultaneously photoexcited by a patterned pump laser for THz diffraction image encoding, which is then detected by a far-field THz single-pixel detector. Right panel: The measured diffraction intensity $I$ and patterns $P$ are utilized to reconstruct a low-quality THz diffraction image $\hat{O}'_d$ through eq 5. A lightweight untrained neural network $f_\theta$ (see Figure S1 in Supporting Information,) simulates the physical processes of diffraction and SPI. Using the low-quality image $\hat{O}'_d$ as input, the network generates a high-quality, high-resolution output image $\hat{O}_0$. This output undergoes numerical propagation via the ASP model to produce a diffraction image $\hat{O}_d$, followed by simulated single-pixel measurements with patterns $P$ to generate estimated 1D intensity $\hat{I}$. The network parameters $\theta$ are optimized by minimizing the loss between $\hat{I}$ and $I$. The dashed arrows

of the images $\hat{O}_0$ and $\hat{O}_d$ illustrate the iterative optimization process, while the black and blue dashed arrows represent data-flow pathways. The solid arrows denote mathematical operations labeled on the arrows (see METHODS for details).

## RESULTS AND DISCUSSION

### Experimental Setup.

As illustrated in Figure 2a, the experimental setup for the untrained TSPIDL system is constructed using a CW 0.36-THz source ($\lambda$ = 833.3 μm, output power of 1 mW), which includes a microwave signal generator (SMB 100A) and THz frequency multiplier modules (VDI WR9.0 M-SGX, WR2.8 × 3). The emitted THz beam is first collimated with a 90° off-axis parabolic (OAP) mirror and directed onto the object. A high-resistivity silicon wafer ( > 10000 Ω·cm) with a 500-μm thickness and a 300-nm-thick $SiO_2$ layer on top is used as the THz photomodulator. The scattered field from the object propagates through free space for a certain distance before reaching the passivated silicon modulator. Subsequently, a CW 532-nm beam is patterned using a digital micromirror device (DMD), focused by Lens3, and projected through the aperture of 90° OAP1 mirror onto the passivated silicon wafer to generate high-energy charge carriers. The spatial distribution of these photogenerated carriers then encodes the THz image. Finally, the transmitted THz signal is focused by OAP2 mirror and detected by an AlGaN/GaN HEMT detector (Suzhou XinZhen Electronic Technology). The data is collected and transmitted to a computer for postprocessing via a data acquisition card (National Instruments USB-6361).

### Spatial Encoding of the THz Wave.

The spatial modulation of the THz beam is a key component of the THz untrained TSPIDL system, where we exploit photo-induced conductivity changes in silicon to control the THz wave.[17,46] Through chemical passivation, a $SiO_2$ dielectric layer is applied to the high-resistivity silicon surface to terminate dangling bonds in the lattice.[47] This process reduces surface trap states and lowers the surface recombination rate of photogenerated carriers. Figure 2b illustrates the modulation geometry, where a 532-nm laser normally illuminates the passivated side of the high-resistivity silicon wafer, forming a thin conductive layer of free charge carriers. The thickness of the thin conductive layer (~8 μm) is determined by the penetration depth of the 532-nm exciting light into the silicon wafer.[48] Due to the increase in free charge carriers, the dielectric function, $\varepsilon(\omega)$, of silicon can be modeled using the Drude model,[17,23,46] which is defined as:

$$\varepsilon(\omega) = \varepsilon_\infty + \frac{i\sigma(\omega)}{\omega\varepsilon_0} = \varepsilon_\infty - \frac{\omega_p^2}{\omega(\omega + i/\tau_d)} \tag{9}$$

where $\sigma(\omega)$ is the complex conductivity, $\omega$ is the frequency of the THz wave, $\varepsilon_\infty$ is the background permittivity of the silicon, $\varepsilon_0$ is the dielectric constant in vacuum, $\tau_d$ is the Drude damping time, and $\omega_p$ is the plasma frequency associated with the photocarrier concentration. In its normal state, the $\omega_p$ of silicon lies outside the THz range, leading to a dielectric response of the silicon wafer. However, under intense photoexcitation, the increased photocarrier concentration elevates the $\omega_p$ into the THz domain, effectively transforming the material response into that of a conductor. This dielectric-to-conductor transition enables modulation of the transmission intensity of the THz wave (Figure 2b).

As the pump laser power increases, the photocarrier density rises rapidly, resulting in an increase in conductivity. This, in turn, leads to stronger THz wave attenuation and a decrease in the transmitted THz amplitude. In addition, upon 532-nm photoexcitation, the reflectivity at the air/$SiO_2$/Si interface varies with the thickness of the $SiO_2$ layer (Figure S3, Supporting

Information), thereby affecting the effective utilization of the pump power for generating high-energy charge carriers in the silicon wafer. A 270-nm-thick $SiO_2$ layer yields the lowest reflectivity of the excitation light, representing the optimal thickness for maximizing pump absorption. As shown in Figure 2c, the modulator efficiency is characterized by its transmission modulation factor, calculated as $(E–E')/E$, where $E$ represents the THz intensity transmitted through the unilluminated wafer, and $E–E'$ denotes the differential intensity change induced by photoexcitation. The modulation depth curve was obtained by systematically measuring the THz transmission changes while varying the pump power density. The saturated absorption is observed under a pump illumination of 0.9 $W/cm^2$, with the corresponding initial power being 1.2 $W/cm^2$. Notably, under equivalent pump fluence, it has been demonstrated that reducing the silicon wafer thickness leads to decreased modulation depth due to carrier diffusion effects.[22,49] Due to the pump beam being patterned by the DMD, different photocarrier distributions are "imprinted" onto the silicon surface, enabling spatial modulation of the THz wave. Specifically, the area illuminated by the pump laser becomes conductive, encoding the THz beam as state "0" (low transmission intensity), while the other areas remain transparent, encoding the THz beam as state "1". As depicted in Figure 2d, a Walsh-Hadamard matrix encoding scheme is adopted, where the matrix is an orthogonal square matrix consisting of "+1" and "−1" elements. However, DMD can only generate Hadamard-like mask with the values of "+1" and "0". As shown in Figure 2e, a Hadamard pattern $P_i(x, y)$ with [+1, −1] values can be realized through differential measurement $P_i^{-}(x, y) = P_i^{+}(x, y) – P_i^{-}(x, y)$. Here, $P_i^{+}(x, y)$ is a positive mask with [+1, 0] values and $P_i^{-}(x, y)$ is a negative mask with [0, +1] values. Thus, the detector readout is obtained through differential measurements, which effectively suppress strong background noise $\eta$ in the actual measurements (see eq 4), and is represented as:

$$I_i = I_i^{+} - I_i^{-} = \iint O_d\left(x,y\right)\left(P_i^{+}\left(x,y\right) - P_i^{-}\left(x,y\right)\right)dxdy \quad (10)$$

For image reconstruction with $N$ pixels using $M$ orthogonal Hadamard patterns $P_i(x, y)$, the sampling ratio is defined as $SR = M / N$. The THz image can then be reconstructed using $I = \{I_i \mid i = 1, 2, \ldots, M\}$ and the corresponding mask patterns $P = \{P_i \mid i = 1, 2, 3, \ldots, M\}$ with the untrained TSPIDL as previously described in METHODS.

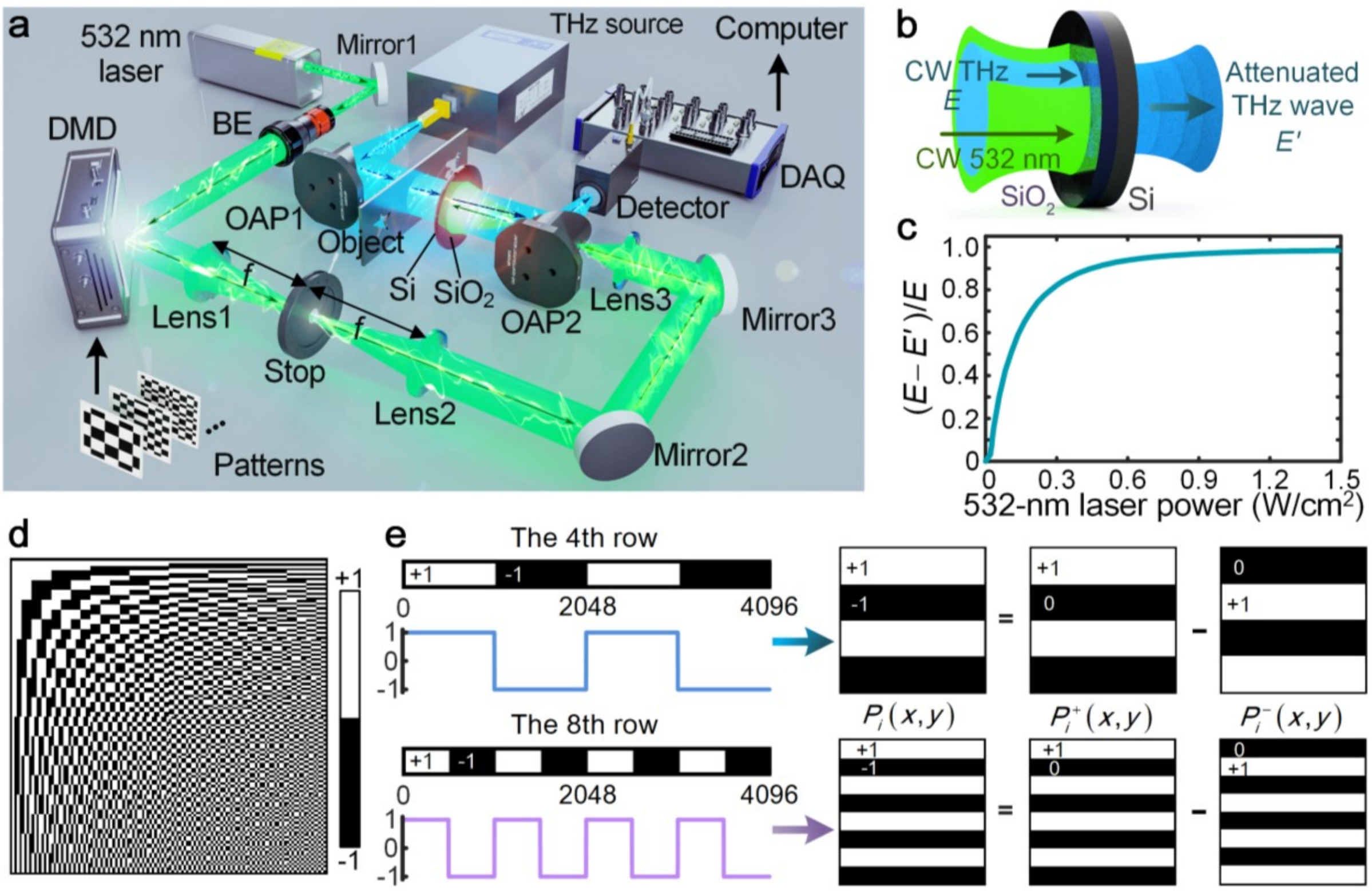

**Figure 2**. (a) Schematic of the CW THz untrained TSPIDL system. DMD: digital micromirror device that provides amplitude modulation with “0” and “1”. BE: beam splitter. Lens1–3: D = 25.4 mm, $f_1 = f_2 = 100$ mm, $f_3 = 30$ mm. Lens1 focuses the DMD-reflected light while blocking diffracted light through an aperture stop, Lens2 is collimating lens and Lens3 is a focusing lens. OAP1–OAP2: 90° off-axis parabolic mirror, D = 50.8 mm, RFL = 152.4 mm, PFL = 76.2 mm. (b) Passivated silicon modulation geometry. The pump laser is normally incident on the passivation surface, while the attenuated THz wave exits from the unpassivated side. (c) THz modulation depth as a function of pump power densities. The transmission efficiency of the optical system at a wavelength of 532 nm is approximately 75% (with a loss of 25%), mainly due to the diffraction efficiency and the surface reflectivity of the micromirror of DMD micromirror. (d) A typical 64 × 64-order Walsh-Hadamard matrix used to encode the THz wave. (e) The 4th and 8th row vectors of (d). Right panel: differential Hadamard patterns from (d).

**Untrained TSPIDL SPI without ASP.**

To evaluate the generalization capability and robustness against noise of the untrained TSPIDL framework for THz SPI, the imaging performance of the method was experimentally studied without considering the ASP model. Imaging was conducted on objects at forward propagation distances of $d_{FP}$ = 0.5 mm and $d_{FP}$ = 4 mm from the back surface of the silicon wafer, where the former corresponds to high-quality images with the majority of near-field components recorded, while the latter represents typical images that are significantly blurred by diffraction and environmental noise. As illustrated in Figure 3a,b, two hollow metallic masks with intricate structures were positioned directly on the unpassivated layer of the silicon wafer, corresponding to a forward propagation distance $d_{FP}$ = 0.5 mm. THz diffraction images of the objects were undersampled using differential Hadamard masks, and the untrained TSPIDL model was employed to reconstruct their THz images from the undersampled data. In this setup, the mask switching rate was set to 1 kHz, and the reconstructed images had a pixel resolution of 64 × 64 within a field-of-view of 10 mm × 10 mm. As shown in Figure 3a2–a6 and Figure 3b2–b6, the reconstructed images exhibit minimal background noise across various examples. Additionally, we compared reconstructed images using the untrained TSPIDL with positive, negative, and differential Hadamard masks under the same *SR* (Figure S4, Supporting Information). The results clearly show that differential masks yield images with higher SNR, while the use of only positive or negative masks introduces pronounced artifacts. To quantitatively evaluate the visual appearance and quality of the reconstructed images, the structural similarity index (SSIM),[50] which ranges from 0 to 1, is employed as a performance metric. The fully sampled THz images reconstructed using HSPI,[17] a model-driven approach that provides a mathematically rigorous and closed-form solution for image reconstruction, serve as reference images (Figure 3a1,b1) for SSIM calculation under varying *SR*. Higher SSIM values, as displayed at the bottom of Figure 3a2–a6 and Figure 3b2–b6, indicate a closer resemblance between the reconstructed and reference images. A similar upward trend in SSIM values is observed in the simulation results (see Figure S5 in Supporting Information). The experimental results demonstrate that the proposed method successfully reconstructs THz images of good quality, with the SSIM reaching approximately 0.76 and 0.71 at *SR* = 3.125% (Figure 3a3,b3). Even at a low *SR* = 1.5625%, the SSIM can reach 0.67 and the reconstructed images are acceptable (Figure 3a1,b1). In previous studies, a *SR* of approximately 80% was required to reconstruct images of similar quality (SSIM = 0.73) using the HSPI technique.[17] The superior performance of the untrained TSPIDL is attributed to its robust learning capabilities, efficient optimization algorithms, multi-layer feature extraction, complex nonlinear mapping abilities, and remarkable adaptability and generalization. These features enable the untrained TSPIDL to effectively reconstruct high-quality images from significantly lower *SR*s, positioning it as a powerful and efficient inverse method.

To further evaluate the robust reconstruction capabilities of our proposed method for diffracted THz images, a metallic mask with a hollow star outline (Figure 3b) was placed at a distance of $d_{FP}$ = 4 mm from the surface of the passivated layer. A comparative study was conducted under identical conditions between the untrained TSPIDL algorithm and three used SPI reconstruction methods: HSPI, total variation-based compressed sensing (CS-TV),[24] and a dataset-trained TSPIDL model. The setup of the trained TSPIDL is provided in Figure S1 (Supporting

Information), and the network architecture is identical to that shown in Figure 1. The reconstructed images under two different *SR*s are depicted in Figure 3c1–f2. From a visual perspective, the deep-learning-based untrained TSPIDL method (Figure 3f1,f2) demonstrates significantly greater robustness to noise compared to traditional methods (Figure 3c,d). Specifically, the images reconstructed using untrained TSPIDL (Figure 3f1,f2) exhibit high contrast and clarity, whereas those reconstructed by HSPI (Figure 3c1,c2) and CS-TV (Figure 3d1,d2)) suffer from substantial noise and blurring.

Notably, at a low *SR* of 3.125%, the HSPI method produces distorted results (Figure 3c1), while the trained TSPIDL method fails to reconstruct high-fidelity images across all *SR*s (Figure 3e1,e2). This limitation stems from the heavy dependence of the trained TSPIDL on its training dataset and specific imaging tasks, a characteristic inherent to supervised learning approaches commonly employed in THz SPI[31,33]. When reconstructing images of objects not encountered during training and in the presence of severe measurement noise, the generalization limitations of the trained TSPIDL become evident, resulting in distorted images. In contrast, the untrained TSPIDL method exhibits strong generalization capabilities, comparable to traditional model-driven SPI reconstruction methods like HSPI and CS-TV (Figure 3c2,d2), while significantly outperforming them in terms of SNR at low *SR* (Figure 3f1). Quantitative evaluations presented in Figure 3g reveal that the untrained TSPIDL achieves approximately a two-fold improvement in SNR under noisy conditions compared to state-of-the-art methods.

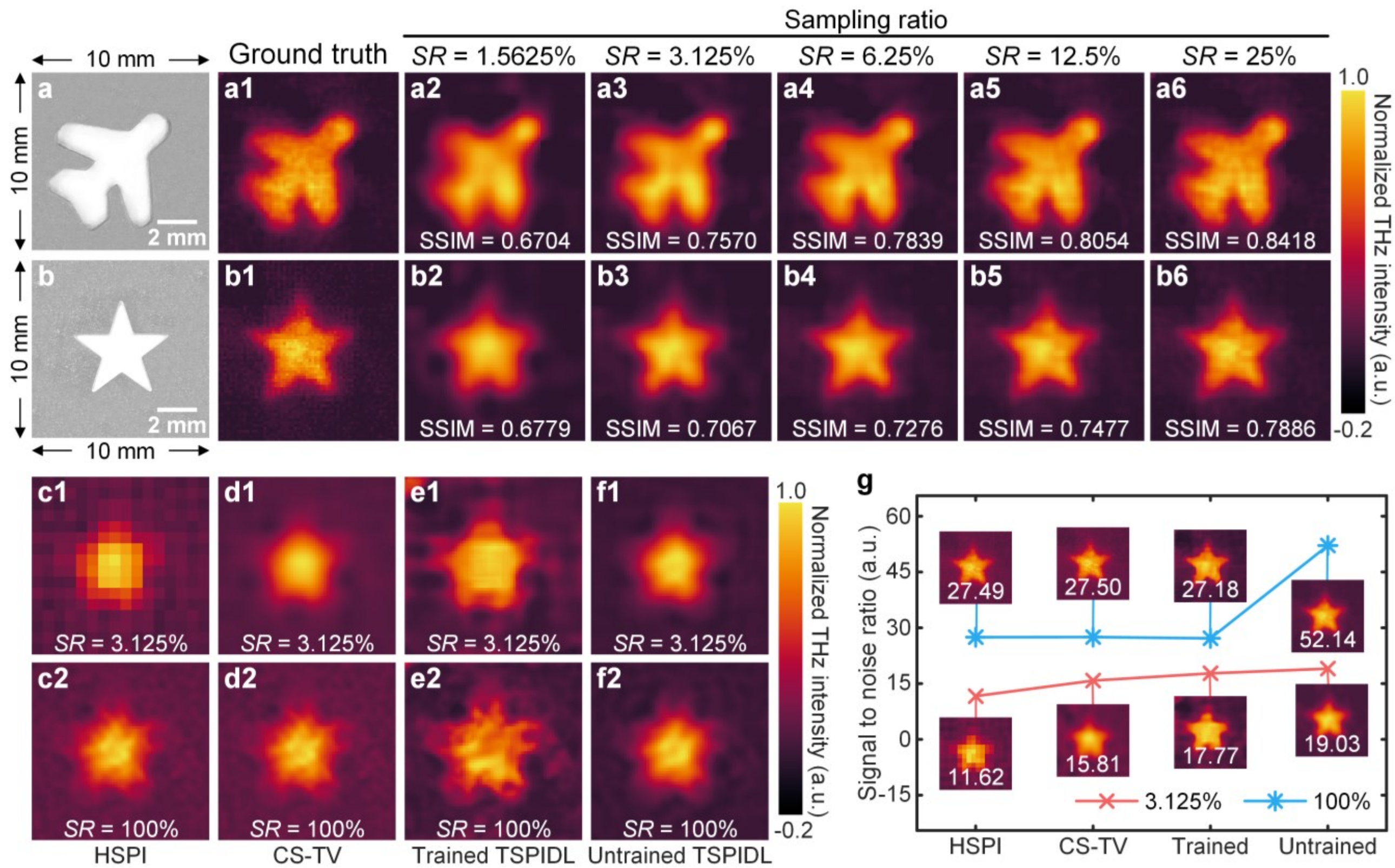

**Figure 3**. Results of the THz SPI without the ASP model. (a) and (b) are the optical images of different objects. (a1) and (b1) are the results of the HSPI full sampling reconstruction. (a2–a6) and (b3–b6) are the THz images ($d_{FP}$ = 0.5 mm) reconstructed by the untrained TSPIDL under different *SR*. In comparison with HSPI (c1–c2), CS-TV (d1–d2), and trained TSPIDL (e1–e2), our method (f1–f2) demonstrates state-of-the-art imaging results ($d$ = 4 mm) under *SR* = 3.125% and *SR* = 100%. The SNRs of the images corresponding to (b) at $d$ = 0 mm, reconstructed by different methods, are shown in (g). The details in calculation of SNR are given in Figure S6 (Supporting Information). The untrained TSPIDL model is optimized for 200 iterations. The FOV for all THz images is 10 mm × 10 mm with 64 × 64 pixels.

**Backpropagation SPI.**

Although the untrained TSPIDL can reconstruct the THz image of the object at low *SR*, the image remains blurred by the diffraction effect because the THz near-fields cannot be fully recorded. Even when the object is placed close to the front surface of the silicon wafer (Figure 1, $d_{FP}$ = 0.5 mm), the finite thickness of the wafer (500 μm) limits the sampling of THz waves, preventing full access to the near-field region ($d_{FP}$ = 500 μm $<$ $\lambda$ = 833.3 μm). Moreover, as the forward propagation $d_{FP}$ increases, the near-fields decay exponentially, giving rise to more blurred images (see Figure S7 in Supporting Information). To overcome the diffraction effect and reconstruct an undistorted image, we propose embedding the ASP model into the untrained neural network (eq 4–6). To clearly demonstrate the sub-diffraction imaging capability, we prepare three metallic masks, each with three air slits. The slit widths are 1217 μm, 884 μm, and 920 μm, respectively, while the separations between adjacent slits are 806 μm (Figure 4a), 405 μm (Figure 4b), and 118 μm (Figure 4c), corresponding to ~$\lambda$, $\lambda/2$, and $\lambda/7$, respectively.

The objects were placed in close contact with the unpassivated silicon wafer at a forward propagation distance of $d_{FP}$ = 0.5 mm. To resolve the minimum slit width of 118 μm (~$\lambda/7$) shown in Figure 4c, a 256 × 256-order Walsh-Hadamard matrix is used to reconstruct each pixel (~41 μm) within the imaging field of view (FOV) of 10.5 mm × 10.5 mm. The scattering field resulting from the interaction between the THz field and the sub-wavelength metallic masks can be captured at a near-field distance within one wavelength ($\lambda$ = 833.3 μm). This includes weak evanescent field components, enabling THz imaging with sub-wavelength resolutions of 806 μm and 405 μm, as shown in the imaging results in Figure 4d,e. When the object size is reduced to $\lambda/7$, as shown in Figure 4f, the resolution deteriorates. This resolution degradation is attributed to the exponential decay of the near-field with propagation distance (see eq 3). As discussed in references,[23,24,26] reducing the modulator thickness to a value much smaller than the object resolution is crucial for enhancing the final resolution. However, in practical imaging scenarios, a certain propagation distance between the modulator and the object exists, and the inherent thickness of the silicon wafer cannot be neglected. As a result, the THz images inevitably suffer from resolution errors due to diffraction effects. Additionally, we notice that a similar work recently demonstrated a resolution of $\lambda_{0.14\text{THz}}/7$ using a 500-μm-thick polymer silicon combined with high-*SR* compressed sensing algorithms. However, it should be emphasized that our approach provides a general framework for further enhancing spatial resolution while maintaining the same silicon wafer thickness.

To achieve sub-diffraction imaging with a spatial resolution of $\lambda/7$, we integrated the ASP model into the output layer of the untrained TSPIDL, enabling it to simulate the diffraction process of THz waves propagating from the front surface to the back surface of the silicon wafer, with the wafer thickness serving as the propagation distance (for further details, see Methods and Figure 1). Specifically, the proposed method was applied to reconstruct the object (Figure 4c) placed in close contact with the back surface of the silicon wafer (at the plane $z = 0$). By accurately inputting the backpropagation distance $d_{BP}$ into eq 2 and eq 3, the untrained TSPIDL, guided by the ASP model, effectively reconstructs the THz image backpropagated to plane $z = -d_{BP}$, relative to the back surface of the silicon. For the object (Figure 4c) placed on the front surface of the silicon wafer (corresponding to a forward propagation distance of $d_{FP} = 0.5$ mm from the back surface), we retrieved a series of images employing the untrained TSPIDL algorithm with varying backpropagation distances $d_{BP}$ of 0.1 mm, 0.3 mm, 0.5 mm, and 1.0 mm. The retrieval results are illustrated in Figure 4f1−f4, clearly demonstrating that correlation between backpropagation distance and spatial resolution.

Specifically, for a backpropagation distance of $d_{BP}$ = 0.1 mm, as shown in Figure 4f1, it is evident that the subwavelength features remain heavily influenced by the THz diffraction field, resulting in a significantly blurred image. As the backpropagation distance increases to $d_{BP}$ = 0.3 mm, the reconstructed THz image begins to recover some of the three-slit features, though distortions persist (Figure 4f2). Only at a backpropagation distance of $d_{BP}$ = 0.5 mm—equal to the thickness of the silicon wafer—is a relatively clear and sharp three-slit THz image obtained (Figure 4f3). This result demonstrates that, even with the presence of a relatively thick silicon wafer (500 µm), the diffracted 0.36-THz image can recover missing details and achieve a sub-diffraction spatial resolution of 118 µm ($\lambda$/7). However, when the backpropagation distance is increased to $d_{BP}$ = 1.0 mm, the retrieved image becomes distorted again (Figure 4f4). For a clearer comparison, Figure 4g quantitatively evaluates the resolution of the retrieved images at backpropagation distances of $d_{BP}$ = 0.1 mm and $d_{BP}$ = 0.5 mm against the optical image of the original object. The overall trend reveals that as the backpropagation distance approaches the thickness of the silicon wafer, the reconstructed images become increasingly distinct. This phenomenon resembles the refocusing of distorted THz images (Figure 4f1) onto a focal plane (Figure 4f3), with distortions reemerging (Figure 4f4) once the image deviates from this focal plane. These results unequivocally demonstrate that embedding the ASP model into the untrained TSPIDL enables the reversal of the THz diffraction field, thereby achieving sub-diffraction THz imaging. It should be noted that our proposed algorithm relies on prior knowledge of the object position, specifically the forward propagation distance $d_{\mathrm{FP}}$, to achieve high-quality image reconstruction. In the absence of this information, artifacts can appear in the reconstructed images. For instance, when the backpropagation distance $d_{\mathrm{BP}}$ = 1.0 mm deviates from the actual forward propagation distance $d_{\mathrm{FP}}$ = 0.5 mm, artifacts emerge—such as the appearance of four apertures instead of the original three (Figure 4f4). Addressing this limitation will require further investigation into adaptive algorithms that can dynamically adjust based on feedback from the reconstructed output. Although this represents a promising direction, it falls beyond the scope of the present work and will be pursued in future studies.

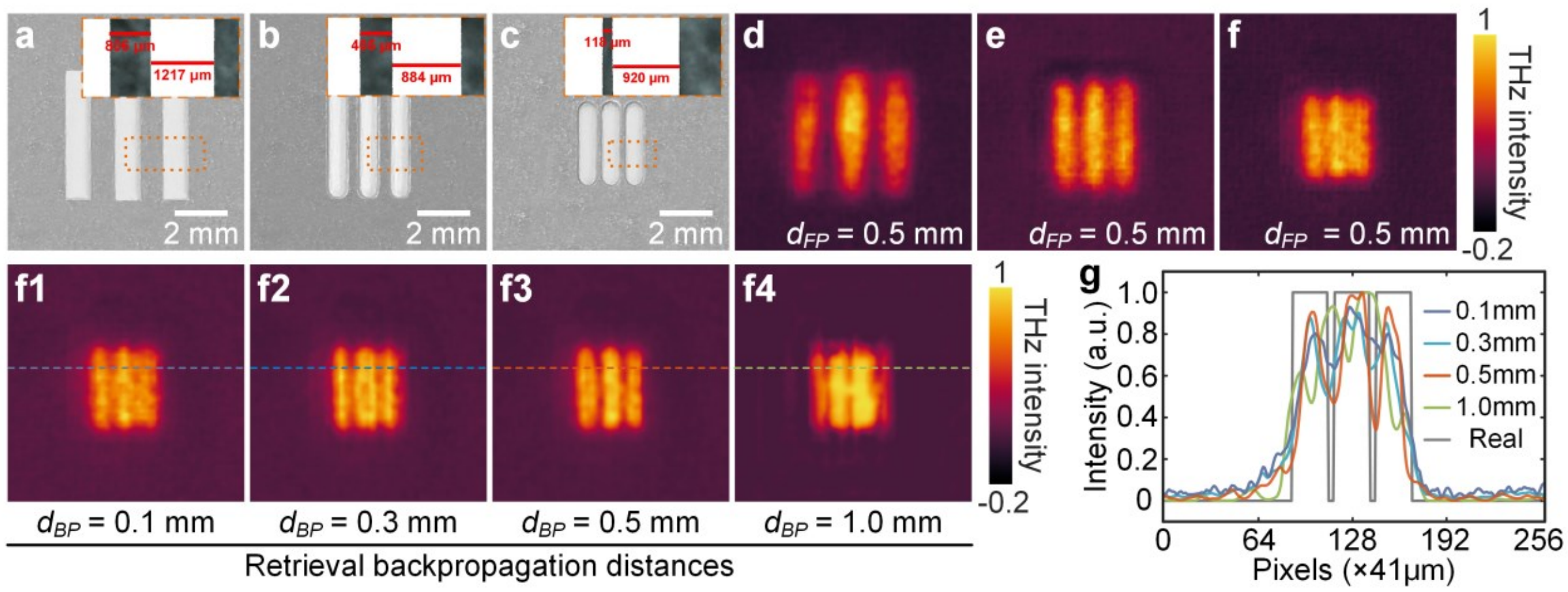

**Figure 4**. Sub-diffraction imaging using the untrained TSPIDL embedded with the ASP model. (a–c) are the optical images of different metallic objects with varying slit widths and adjacent widths (insets are the optical images under 5× magnification). (d–f) are the THz diffraction images ($d_{FP}$ = 0.5 mm) reconstructed by untrained TSPIDL. (f1–f4) are the sub-diffraction THz images reconstructed by the proposed method at varying backpropagation distances($d_{BP}$ = 0.1 mm, 0.3 mm, 0.5 mm and 1.0 mm) when $d_{FP}$ = 0.5 mm. (g) Cross-sectional intensity along the dash line shown in (f1–f4). The untrained TSPIDL model is optimized for 200 iterations. The FOV for all THz images is 10.5 mm × 10.5 mm with 256 × 256 pixels, and under the sampling ratio of $SR$ = 3.125%.

Another important aspect to investigate is the applicability of the untrained TSPIDL for imaging objects located at relatively large distances from the silicon plane, i.e., without sampling the near-field components. As previously shown in Figure 3f2, the THz image of a far-field object at $d_{FP}$ = 4 mm becomes significantly blurred by the diffraction effect. This is due to the rapid decay of the THz near-fields (eq 3) to nearly zero in the far-field region. To further validate the ability of the proposed method to reconstruct THz images at varying distances from the silicon plane, we placed metallic masks with hollow plane and star outlines (Figure 3a,b) at distances of $d_{FP}$ = 2 mm, $d_{FP}$ = 4 mm, and $d_{FP}$ = 6 mm from the back surface of the silicon wafer. The corresponding THz diffraction images at these three distances were then reconstructed using the untrained TSPIDL method, without the ASP model. At $d_{FP}$ = 2 mm (Figure 5a,b), the outlines of the airplane and star-shaped masks are clearly discernible, with only minor loss of detail at the hollow edges. However, at $d_{FP}$ = 4 mm (Figure 5c,d) and $d_{FP}$ = 6 mm (Figure 5e,f), the diffraction effects from the object significantly impact THz transmission, as discussed earlier. As a result, subwavelength details rapidly degrade during propagation, making them nearly indistinguishable in the reconstructed images. In contrast, by embedding the untrained TSPIDL with the ASP model to perform backpropagation on the diffracted images, we can successfully refocus these diffraction patterns generated on different planes $z = d_{FP}$ back to the planes $z = 0$. As shown in Figure 5g–l, when backpropagation imaging is applied at the corresponding distances, most of the object details are successfully recovered. However, as the propagation distance increases in free space, image retrieval becomes increasingly challenging. It is evident from Figure 5g–l that the image quality, including details and contrast, deteriorates compared to those shown in Figure 3a6 and Figure 3b6. This can be attributed to the fact that as the propagation distance grows, higher spatial frequencies are deflected at larger angles relative to the optical axis. The limited FOV of the imaging system results in the loss of high-frequency spatial information in the diffraction field. Therefore, for accurate reconstruction of the backpropagation images (Figure 4f3 and Figure 5g–l), consideration must be given to sufficiently high detector sensitivity and modulation depth of the THz waves, precise propagation distance and FOV size, as well as the noise robustness of the optical path with an adequately large imaging FOV. On our computational platform (Intel Xeon Platinum 8336C CPU and NVIDIA GeForce RTX 4090 GPU), the untrained TSPIDL algorithm requires approximately 3 minutes to reconstruct a fully-sampled 64 × 64 image. In comparison, HSPI, CS-TV, and the trained TSPIDL demonstrate significantly faster reconstruction times of 0.004 s, 8 s, and 0.08 s, respectively. With a mask modulation rate of 1 kHz, the hardware acquisition time for fully-sampled measurements is approximately 9 seconds. This reveals that the software reconstruction time (180 s) of the untrained TSPIDL substantially exceeds its hardware acquisition time by a factor of 20, indicating that the overall system throughput is currently bottlenecked by computational performance rather than data acquisition speed. To improve computational efficiency of the untrained TSPIDL, several strategies can be considered, such as optimizing neural network architectures, improving the initialization strategies, and leveraging high-performance computing platforms.

It should be noted that while our approach combining backpropagation and deep neural networks enables subwavelength imaging, the resolution of subwavelength features remains relatively weak. Such limitation can be attributed to both hardware-related and algorithmic factors. On the hardware side, we use silicon wafers to modulate THz waves via photogenerated carriers; however, lateral carrier diffusion within the silicon may degrade imaging quality. On the algorithmic side, we acknowledge that the proposed method may introduce artifacts during angular spectrum propagation, primarily due to the inherently highly compressed nature of single-pixel imaging signals. Moreover, the reconstruction process relies on prior knowledge of the object's position (i.e., $d_{FP}$) to produce high-quality images. Despite these limitations, the untrained TSPIDL framework embedded with the ASP model successfully achieves sub-diffraction imaging using relatively thick substrates. This significantly relaxes the stringent experimental requirements imposed by previous approaches and reduces reliance on ultrathin photomodulators.[23,24,26] In future work, alternative modulation strategies that avoid carrier diffusion, such as using materials or devices without carrier diffusion limitations, can be explored to further improve imaging quality.

The untrained TSPIDL model presented in our current study does not require pre-training on any labeled dataset. Instead, it operates through an unsupervised learning strategy with iterative optimization constrained by imaging physics principles, thereby yielding a generalizable model that adapts to the specific imaging task at hand. While our results demonstrate that the untrained TSPIDL model offers a universally applicable approach capable of achieving $\lambda_0/7$ imaging resolution, it is important to acknowledge the inherent limitations imposed by the electromagnetic wavelength. For 0.36 THz radiation with $\lambda_0$ = 833 μm, even a resolution of $\lambda_0/7$ = 118 μm remains relatively coarse. Consequently, reconstructing patterns with sharp features at image boundaries presents significant challenges. To address this limitation, future implementations could potentially achieve enhanced resolution by either reducing the imaging distance $d$ ($d \ll \lambda_0$) or employing higher-frequency THz radiation (corresponding to shorter wavelengths), thereby enabling the reconstruction of sharper features at image boundaries.

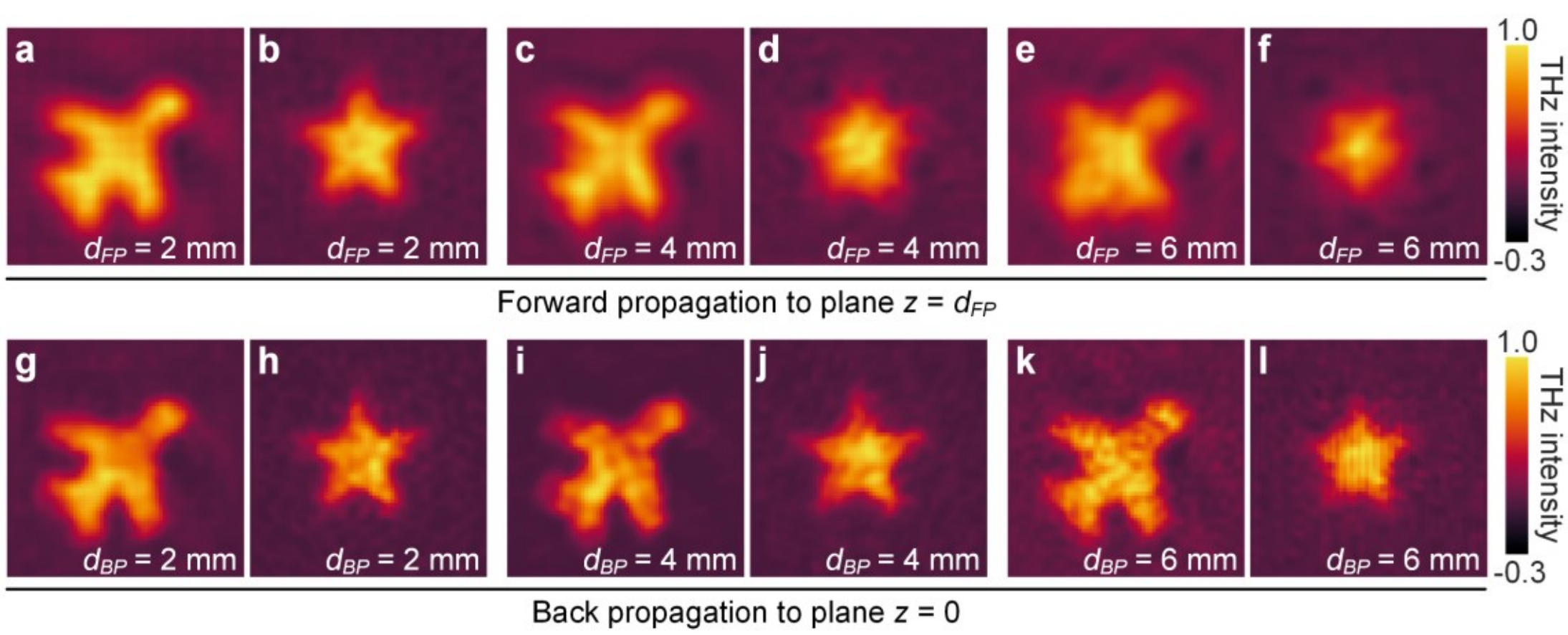

**Figure 5.** Comparison of reconstructed far-field images using the untrained TSPIDL without (a–f) and with (g–l) the ASP model. The metallic masks are placed at $d_{FP}$ = 2 mm (a, b), $d_{FP}$ = 4 mm

(c, d), and $d_{FP}$ = 6 mm (e, f) from the back surface of the silicon wafer. The untrained TSPIDL model is optimized for 200 iterations. The FOV for all THz images is 10.5 mm × 10.5 mm with 64 × 64 pixels, and under the sampling ratio of SR = 50%.

## CONCLUSION

In conclusion, we have proposed and experimentally demonstrated sub-diffraction THz backpropagation SPI using an untrained neural network constrained by the ASP model. Our approach effectively couples the diffraction field with THz SPI in a CW monochromatic THz imaging system. A key innovation lies in the backpropagation process, achieved during iterative optimization, in which the amplitude distribution estimated from the network output is used to generate the diffracted THz image at a fixed propagation distance via the ASP model. In the near-field region, we demonstrated a proof-of-principle application where a subwavelength object was placed in close contact with a 500-μm-thick silicon wafer. Our method successfully reverses the effects of diffracted fields from blurred THz images, achieving a spatial resolution of $\lambda/7$ at a low *SR*. Additionally, in the far-field region, the method recovers most object details, enabling the refocusing of THz images at planes with certain distances from the silicon wafer. To our knowledge, this is the first demonstration of reconstructing sub-diffraction THz images in highly undersampled subwavelength objects. The proposed untrained neural network for THz SPI represents a departure from traditional supervised learning methods. The proposed method enhances image SNR by approximately two-fold compared to state-of-the-art techniques and achieves imaging at an ultralow *SR* of 1.5625%. Therefore, our approach provides a general optimization algorithm for further enhancing spatial resolution while maintaining the same silicon wafer thickness. From a hardware perspective, thicker silicon wafers do introduce diffraction effects and lateral carrier diffusion, compromising overall imaging performance, whereas reducing wafer thickness offers improved spatial resolution but sacrifices modulation depth. Future work will focus on this compromise and further explore the relationship between reconstructed spatial resolution and multiple parameters, including silicon wafer thickness, DMD pixel dimensions and its utilization efficiency. Our work not only paves the way for developing THz microscopes capable of non-invasive subwavelength imaging, but also offers a pathway to improve SNR in other near-field and low-signal imaging scenarios.

**ASSOCIATED CONTENT**

**Supporting Information**.

The following files are available free of charge.

Figures S1–S7, details of theoretical modeling, experiment and data analysis. (PDF)

**AUTHOR INFORMATION**

**Corresponding Author**

**Huanjun Chen** — State Key Laboratory of Optoelectronic Materials and Technologies, Guangdong Province Key Laboratory of Display Material and Technology, School of Electronics and Information Technology, Sun Yat-sen University, Guangzhou 510275, China; orcid.org/0000-0003-4699- 009X; Email: chenhj8@mail.sysu.edu.cn

**Shaozhi Deng** — State Key Laboratory of Optoelectronic Materials and Technologies, Guangdong Province Key Laboratory of Display Material and Technology, School of Electronics and Information Technology, Sun Yat-sen University, Guangzhou 510275, China; orcid.org/0000-0003-1830- 2026; Email: stsdsz@mail.sysu.edu.cn

**Authors**

**Yongsheng Zhu** — State Key Laboratory of Optoelectronic Materials and Technologies, Guangdong Province Key Laboratory of Display Material and Technology, School of Electronics and Information Technology, Sun Yat-sen University, Guangzhou 510275, China

**Shaojing Liu** — State Key Laboratory of Optoelectronic Materials and Technologies, Guangdong Province Key Laboratory of Display Material and Technology, School of Electronics and Information Technology, Sun Yat-sen University, Guangzhou 510275, China

**Ximiao Wang** — State Key Laboratory of Optoelectronic Materials and Technologies, Guangdong Province Key Laboratory of Display Material and Technology, School of Electronics and Information Technology, Sun Yat-sen University, Guangzhou 510275, China

**Runli Li** — State Key Laboratory of Optoelectronic Materials and Technologies, Guangdong Province Key Laboratory of Display Material and Technology, School of Electronics and Information Technology, Sun Yat-sen University, Guangzhou 510275, China

**Haili Yang** — State Key Laboratory of Optoelectronic Materials and Technologies, Guangdong Province Key Laboratory of Display Material and Technology, School of Electronics and Information Technology, Sun Yat-sen University, Guangzhou 510275, China

**Jiali Wang** — State Key Laboratory of Optoelectronic Materials and Technologies, Guangdong Province Key Laboratory of Display Material and Technology, School of Electronics and Information Technology, Sun Yat-sen University, Guangzhou 510275, China

**Hongjia Zhu** — State Key Laboratory of Optoelectronic Materials and Technologies, Guangdong Province Key Laboratory of Display Material and Technology, School of Electronics and Information Technology, Sun Yat-sen University, Guangzhou 510275, China

**Yanlin Ke** — State Key Laboratory of Optoelectronic Materials and Technologies, Guangdong Province Key Laboratory of Display Material and Technology, School of Electronics and Information Technology, Sun Yat-sen University, Guangzhou 510275, China

**Ningsheng Xu** — State Key Laboratory of Optoelectronic Materials and Technologies, Guangdong Province Key Laboratory of Display Material and Technology, School of Electronics and Information Technology, Sun Yat-sen University, Guangzhou 510275, China

**Author Contributions**

H. C. and S. D. conceived and designed the experiments. Y. Z. created the experimental setup and programmed all equipment. S. L., X. W., R. L., H. Y., J. W., and H. Z. analyzed the results. Y. Z., H. C., and S. D. wrote the manuscript. H. C., S. D., and N. X. supervised the work. All authors have given approval to the final version of the manuscript.

**Funding Sources**

The National Key Basic Research Program of China (2024YFA1208500, 2024YFA1208501), the National Natural Science Foundation of China (62405381), the Guangdong Basic and Applied Basic Research Foundation (2023A1515011876), the China Postdoctoral Science Foundation (2024M753734), the Postdoctoral Fellowship Program of China Postdoctoral Science Foundation (GZC20242065), the Changjiang Young Scholar Program.

**Notes**

The authors declare no competing financial interest.

## ABBREVIATIONS

THz, Terahertz; SPI, single-pixel imaging; HSPI, Hadamard single-pixel imaging; SR, sampling ratio; CW, continuous-wave; TSPIDL, Terahertz single-pixel imaging based on deep learning; ASP, angular spectrum propagation; CS, compressed sensing; CS-TV, total variation-based compressed sensing; SSIM, structural similarity index; SNR, signal to noise ratio.